\documentstyle[12pt,fps]{article}
\newcommand{\Z}{{\sf Z \!\!\! Z}}

\newcommand{\psibar}{\bar{\psi}}
\newcommand{\Psibar}{\bar{\Psi}}
\makeatletter
\@addtoreset{equation}{section}
\makeatother

\title{A Perturbative Construction of Lattice Chiral Fermions
\footnote{This work is supported in part by funds provided by the U.S.
Department of Energy (D.O.E.) under cooperative research agreement
DE-FC02-94ER40818.}}
\author{W. Bietenholz and U.-J. Wiese \\ \\
Center for Theoretical Physics, \\
Laboratory for Nuclear Science, and Department of Physics \\
Massachusetts Institute of Technology (MIT) \\
Cambridge, Massachusetts 02139, U.S.A. \\ \\
MIT Preprint, CTP 2423 \\ \\}

\begin{document}
\maketitle
\begin{abstract} \normalsize

We perform a renormalization group transformation to construct a lattice theory
of chiral fermions. The field variables of the continuum theory are averaged
over hypercubes to define lattice fields. Integrating out the continuum
variables in perturbation theory we derive a chirally invariant effective
action for the lattice fields. This is consistent with the Nielsen-Ninomiya
theorem because the effective action is nonlocal. We also
construct the axial current on the lattice and we show that
the axial anomaly of the continuum theory is reproduced in the Schwinger model.
This shows that chiral fermions can be regularized on the lattice.

\end{abstract}

\maketitle

\newpage

It is a long-standing problem to construct a lattice regularization of chiral
gauge theories such as the Standard model. Naive discretization of the
continuum action leads to a multiplication of fermion species --- known as the
fermion doubling problem. To remove the doubler fermions Wilson has introduced
a chiral symmetry breaking term \cite{Wil74}. In a chiral gauge theory this
term is problematic because it breaks gauge invariance. Moreover, the
Nielsen-Ninomiya theorem \cite{Nie81} excludes a chirally invariant solution
of the doubling problem assuming hermiticity, locality, and lattice translation
invariance of the fermionic action. An early proposal to circumvent the
doubling problem was made by the SLAC group using a nonlocal action
\cite{Dre76}. That nonlocality manifests itself in a finite discontinuity of
the inverse fermion propagator in momentum space.
However, Karsten and Smit demonstrated that
proposals of this kind fail to reproduce Lorentz invariance in the continuum
limit \cite{Kar78}. In a refined proposal Rebbi introduced nonlocality
by poles instead of finite discontinuities \cite{Reb87}. However, as soon as
gauge
interactions are switched on the doublers return as spurious ghost states
\cite{Bod87,Pel88}. In particular, they cancel the anomaly of the
original fermion and render the theory vector-like. Currently discussed
proposals to solve the doubling problem include domain wall fermions
\cite{Kap90}, gauge fixing approaches \cite{Rom90}, Pauli-Villars/SLAC fermions
\cite{Sla90}, as well as continuum
fermions coupled to an interpolated lattice gauge field \cite{Goe90}.

Here we construct a manifestly gauge invariant
lattice theory of chiral fermions using renormalization
group concepts. Iterating renormalization group transformations a recent study
found a line of fixed points for free Wilson fermions \cite{Wie93}. The fixed
point action at the line's endpoint is nonlocal and corresponds to a theory of
free chiral fermions. In contrast to SLAC and Rebbi fermions here the
nonlocality arises naturally by integrating out the high
momentum modes of the fermion field. As for Rebbi fermions the nonlocality
manifests itself by poles in the inverse propagator. We investigate the
vicinity of the
nonlocal fixed point by switching on a small gauge coupling. Using perturbation
theory we construct the action as well as the axial current,
and we demonstrate that the correct axial anomaly is reproduced in the
Schwinger
model. We add a few remarks on possible nonperturbative implications of our
results.

First we rederive the nonlocal
free fixed point action by blocking directly out of the
continuum. For this purpose we average the continuum fermion fields $\psibar$,
$\psi$ over unit hypercubes $c_x$ centered at $x$ to define lattice variables
\begin{equation}
\Psi_x = \int_{c_x} d^dy \psi(y), \ \ \Psibar_x = \int_{c_x} d^dy \psibar(y).
\end{equation}
This is equivalent to a lattice renormalization group transformation with a
blocking factor that goes to infinity.
The Euclidean effective action $S[\Psibar,\Psi]$ for the lattice variables is
obtained by integrating out the continuum fields
\begin{equation}
\label{effact}
\exp(- S[\Psibar,\Psi]) = \int {\cal D}\psibar {\cal D}\psi
\exp(- s[\psibar,\psi]) \prod_x
\delta(\Psibar_x - \int_{c_x} d^dy \psibar(y))
\delta(\Psi_x - \int_{c_x} d^dy \psi(y)),
\end{equation}
where $s[\psibar,\psi]$ is the free massless fermion action in the continuum.
Eq.(\ref{effact}) represents a chirally
covariant renormalization group transformation. In momentum space
the lattice field is given by
\begin{equation}
\label{freeRGT}
\Psi(p) = \sum_{l \in \Z^d} \Pi(p + 2 \pi l) \psi(p + 2 \pi l), \ \
\Pi(p) = \prod_{\mu = 1}^d \frac{\hat{p}_\mu}{p_\mu}, \ \ \hat{p}_\mu =
2 \sin(p_\mu/2),
\end{equation}
and analogously for $\Psibar$.
This leads to the effective action
\begin{eqnarray}
&&S[\Psibar,\Psi]=\frac{1}{(2 \pi)^d}
\int_B d^dp \Psibar(-p) \Delta^f(p)^{-1} \Psi(p), \nonumber \\
&&\Delta^f(p)=\sum_{l \in \Z^d} [i \gamma_\mu (p_\mu + 2 \pi l_\mu)]^{-1}
\Pi^2(p + 2 \pi l),
\end{eqnarray}
where $B = ]-\pi,\pi]^d$ is the Brillouin zone and $\Delta^f$
is the lattice fermion propagator. The technique of blocking out of the
continuum was first used by Wilson \cite{Wil76} and applied to fermions by
Mack's students Cronj\"ager and Mai \cite{Cro85}. It proves to be extremely
useful for our purposes.

Similarly, we consider an Abelian gauge field $a_\mu$ in the continuum and
construct from it a noncompact lattice gauge field
\begin{equation}
A_{\mu,x} = \int_{c_{x - \hat{\mu}/2}} d^dy (1 + y_\mu - x_\mu) a_\mu(y)
+ \int_{c_{x + \hat{\mu}/2}} d^dy (1 - y_\mu + x_\mu) a_\mu(y),
\end{equation}
where $x$ now refers to the center of a link, i.e. $x_\mu$ is a half-odd
integer. This expression is gauge covariant,
i.e. a gauge transformation $a_\mu' = a_\mu + \partial_\mu \varphi$ in the
continuum induces a lattice gauge transformation $A_{\mu,x}' = A_{\mu,x}
+ \Phi_{x + \hat{\mu}/2} - \Phi_{x - \hat{\mu}/2}$ where $\Phi_x =
\int_{c_x} d^dy \varphi(y)$. In momentum space this implies
\begin{equation}
\label{gaugeRGT}
A_\mu(p) = \sum_{l \in \Z^d} \Pi_\mu(p + 2 \pi l) (-1)^{l_\mu}
a_\mu(p + 2 \pi l), \ \ \Pi_\mu(p) = \frac{\hat{p}_\mu}{p_\mu} \Pi(p).
\end{equation}
Note that $A_\mu(p)$ is antiperiodic over the Brillouin zone in the
$\mu$-direction.
Integrating out the continuum gauge field we obtain the effective action
\begin{eqnarray}
&&S[A]=\frac{1}{(2 \pi)^d} \int_B d^dp A_\mu(-p) \Delta_{\mu\nu}^g(p)^{-1}
A_\nu(p), \nonumber \\
&&\Delta_{\mu\nu}^g(p)^{-1}=\omega_\mu(p) \delta_{\mu\nu} -
\frac{\hat{p}_\mu \omega_\mu(p) \omega_\nu(p) \hat{p}_\nu}
{\sum_\rho \hat{p}_\rho^2 \omega_\rho(p)}, \nonumber \\
&&\omega_\mu(p)^{-1}=\sum_{l \in \Z^d} (p + 2 \pi l)^{-2}
\Pi_\mu^2(p + 2 \pi l).
\end{eqnarray}
where $\Delta^g$ is the gauge field propagator.

Having treated free fermions and gauge fields we now switch on a small
gauge coupling $e$ in perturbation theory. The block renormalization properties
of lattice fermions in an external gauge fields have also been investigated
by Balaban, O'Carroll, and Schor \cite{Bal91}. To leading order in $e$ the
renormalization group transformation (\ref{freeRGT}) generalizes to
\begin{eqnarray}
\label{RGT}
\Psi(p)&=&\sum_{l \in \Z^d} \Pi(p + 2 \pi l) \psi(p + 2 \pi l) \nonumber \\
&+& e \sum_{l \in \Z^d} \frac{1}{(2 \pi)^d} \int d^dq
K_\mu(p + 2 \pi l,q + 2 \pi l) a_\mu(p - q) \psi(q + 2 \pi l).
\end{eqnarray}
Here $K_\mu(p,q)$ is a regular kernel that specifies the renormalization group
transformation. Gauge covariance imposes the constraint $(p_\mu - q_\mu)
K_\mu(p,q) = \Pi(p-q) \Pi(q) - \Pi(p)$. Again we integrate out the continuum
fields and to order $e$ we obtain an interaction term
\begin{eqnarray}
S[\Psibar,\Psi,A]&=&
\frac{1}{(2 \pi)^{2 d}} \int_{B^2} d^dp d^dq \Psibar(-p) e V_\mu(p,q)
A_\mu(p - q) \Psi(q), \nonumber \\
V_\mu(p,q)&=&\Delta^f(p)^{-1} \Delta_{\mu\nu}^g(p-q)^{-1} \sum_{l,m \in \Z^d}
\frac{\Pi_\nu(p + 2 \pi l - q - 2 \pi m)}{(p + 2 \pi l - q - 2 \pi m)^2}
(-1)^{l_\nu+m_\nu} \nonumber \\
&\times&\Big\{K_\nu(p + 2 \pi l,q + 2 \pi m)
[i \gamma_\sigma(q_\sigma + 2 \pi m_\sigma)]^{-1} \Pi(q + 2 \pi m)
\nonumber \\
&-&K_\nu(-q - 2 \pi m,-p - 2 \pi l)
[i \gamma_\rho(p_\rho + 2 \pi l_\rho)]^{-1} \Pi(p + 2 \pi l) \nonumber \\
&-&[i \gamma_\rho(p_\rho + 2 \pi l_\rho)]^{-1} i \gamma_\mu
[i \gamma_\sigma(q_\sigma + 2 \pi m_\sigma)]^{-1} \Pi(p + 2 \pi l)
\Pi(q + 2 \pi m)\Big\} \nonumber \\ &\times &\Delta^f(q)^{-1}
+\frac{\widehat{(p-q)}_\mu \omega_\mu(p-q)}
{\sum_\rho \widehat{(p-q)}_\rho^2 \omega_\rho(p-q)}
[\Delta^f(q)^{-1} - \Delta^f(p)^{-1}].
\end{eqnarray}
The total effective action $S[\Psibar,\Psi] + S[A] + S[\Psibar,\Psi,A]$ is
gauge invariant to order $e$ because the vertex function $V_\mu$ obeys the
Ward identity
\begin{equation}
\widehat{(p-q)}_\mu V_\mu(p,q) = \Delta^f(q)^{-1} - \Delta^f(p)^{-1}.
\end{equation}

Due to the chiral covariance of the renormalization group
transformation the effective action is chirally invariant. In contrast to
the continuum the lattice fermionic measure is also chirally
invariant. Hence one can formally construct a conserved axial Noether current
$\tilde J_\mu^5$
and argue that the anomaly of the continuum theory disappears on the lattice.
In fact, assuming a general form for $\Delta^f$ and $V_\mu$,
Pelissetto showed that $\langle \hat{p}_\mu \tilde J_\mu^5(p) \rangle_A = 0$
in an arbitrary background lattice gauge field $A$, and he
concluded that the anomaly is cancelled by spurious ghost states \cite{Pel88}.
However, like all objects involved, the axial current $\tilde J_\mu^5$ is
nonlocal and there is no evidence for it being related to the local current
$j_\mu^5(p)$ $=$ $(2 \pi)^{-d} \int d^dq \psibar(p-q) \gamma_\mu \gamma_5
\psi(q)$ of the continuum theory. The latter has the well-known anomaly
$\langle p_\mu j_\mu^5(p) \rangle_a =$
$(e/\pi) \epsilon_{\mu\nu} p_\mu a_\nu(p)$ with the topological charge density
on the right-hand side. In fact, even in the continuum one can
formally construct a gauge invariant axial current
\begin{equation}
\tilde{j}_\mu^5(p) = j_\mu^5(p) - \frac{e}{\pi} \epsilon_{\mu\nu}
[a_{\nu}(p) - p_{\nu} p_{\rho} a_{\rho}(p)/p^{2}]
\end{equation}
which is anomaly free but nonlocal.
Hence one must decide carefully what current to consider on the lattice.
We will now construct a lattice axial current
that is related to the continuum current $j_\mu^5$ by the renormalization
group transformation. In fact, it turns out that this current has the correct
anomaly of the continuum theory.

We construct the lattice axial current
\begin{equation}
J_{\mu,x}^5 = \int_{f_{\mu,x}} d^{d-1}y j_\mu^5(y)
\end{equation}
by integrating the
flux of the continuum current through the boundary $f_{\mu,x}$ between two
adjacent hypercubes $c_{x - \hat{\mu}/2}$ and $c_{x + \hat{\mu}/2}$. Then the
lattice divergence of the axial current is equal to the continuum divergence
integrated over the hypercube $c_x$, i.e.
\begin{equation}
\delta J_x^5 = \sum_\mu (J_{\mu,x + \hat{\mu}/2}^5 -
J_{\mu,x - \hat{\mu}/2}^5) = \int_{c_x} d^dy \partial_\mu j_\mu^5(y).
\end{equation}
In momentum space the current takes the form
\begin{equation}
\label{axcur}
J_\mu^5(p) = \sum_{l \in \Z^d} j_\mu^5(p + 2 \pi l) \Pi_{\neg\mu}(p + 2 \pi l)
(-1)^{l_\mu}, \ \ \Pi_{\neg\mu}(p) = \frac{p_\mu}{\hat{p}_\mu} \Pi(p).
\end{equation}
We couple the current $J_\mu^5$ to an external lattice axial current
$J_\mu^e$ by adding a term $(2 \pi)^{-d} \int_B d^dp J_\mu^e(-p) J_\mu^5(p)$
to the continuum action and again we integrate out the continuum fields.
Taking a functional derivative of the effective action
with respect to $J_\mu^e(-p)$ one identifies
\begin{eqnarray}
\label{latcur}
J_\mu^5(p)&=&\frac{1}{(2 \pi)^d} \int_B d^dq \Psibar(p-q) \Delta^f(p-q)^{-1}
\nonumber \\ &\times&\sum_{l,m \in \Z^d} [i \gamma_\nu
(p_\nu + 2 \pi l_\nu - q_\nu - 2 \pi m_\nu)]^{-1} \gamma_\mu \gamma_5
[i \gamma_\rho (q_\rho + 2 \pi m_\rho)]^{-1} \Delta^f(q)^{-1} \Psi(q)
\nonumber \\
&\times&\Pi(p + 2 \pi l - q - 2 \pi m) \Pi(q + 2 \pi m)
\Pi_{\neg\mu}(p + 2 \pi l) (-1)^{l_\mu} + O(e).
\end{eqnarray}
The $O(e)$ term is essential but for reasons of space it will be displayed
elsewhere \cite{Bie95}.

To show that the current has the correct anomaly
we now turn to the Schwinger model. We compute $\langle J_\mu^5(p) \rangle_A$
by integrating out the lattice fermions. Using eq.(\ref{axcur}) one obtains
\begin{eqnarray}
&&\!\!\!\!\!\!\!\!\!\!\!
\langle J_\mu^5(p) \rangle_A=\int {\cal D}a \sum_{l \in \Z^2}
\langle j_\mu^5(p + 2 \pi) \rangle_a \Pi_{\neg\mu}(p + 2 \pi l) (-1)^{l_\mu}
\exp(- s[a]) \times \nonumber \\
&&\!\!\!\!\!\!\!\!\!\!\!\int {\cal D}D \exp\{\frac{i}{(2 \pi)^2} \int_B d^2q
D_\nu(-q)[A_\nu(q) - \!\!\!\sum_{m \in \Z^2} \Pi_\nu(q + 2 \pi m) (-1)^{m_\nu}
a_\nu(q + 2 \pi m)]\}, \nonumber \\ \,
\end{eqnarray}
where $s[a]$ is the free gauge field action in the continuum, and $D$ is an
auxiliary lattice gauge field that implements eq.(\ref{gaugeRGT}) as a
constraint. Using e.g.
dimensional regularization we reproduce the known continuum
result $\langle j_\mu^5(p) \rangle_a = (e/\pi) (p_\mu/p^2)
\epsilon_{\nu\rho} p_\nu a_\rho(p)$. Integrating out the continuum gauge field
we then find
\begin{eqnarray}
\langle J_\mu^5(p) \rangle_A&=&\frac{e}{\pi} \sum_{l \in \Z^2}
\frac{p_\mu + 2 \pi l_\mu}{(p + 2 \pi l)^2} \epsilon_{\nu\rho}
\frac{p_\nu + 2 \pi l_\nu}{(p + 2 \pi l)^2} \nonumber \\
&\times&\Pi_{\neg\mu}(p + 2 \pi l) (-1)^{l_\mu+l_\rho}
\Pi_\rho(p + 2 \pi l) \Delta_{\rho\sigma}^g(p)^{-1} A_\sigma(p).
\end{eqnarray}
Taking the lattice divergence of the axial current and using
$\hat{p}_\mu \Pi_{\neg\mu}(p + 2 \pi l) (-1)^{l_\mu}$ $=$ $(p_\mu + 2 \pi
l_\mu)
\Pi(p + 2 \pi l)$ one obtains
\begin{equation}
\label{anomaly}
\langle \hat{p}_\mu J_\mu^5(p) \rangle_A=\frac{e}{\pi} \sum_{l \in \Z^2}
\epsilon_{\nu\rho} \frac{p_\nu + 2 \pi l_\nu}{(p + 2 \pi l)^2} \Pi(p + 2 \pi l)
(-1)^{l_\rho} \Pi_\rho(p + 2 \pi l) \Delta_{\rho\sigma}^g(p)^{-1} A_\sigma(p).
\end{equation}
Now we construct the lattice topological charge density
\begin{equation}
Q_x = \frac{e}{\pi} \int_{c_x} d^2y \epsilon_{\mu\nu} \partial_\mu a_\nu(y).
\end{equation}
Treating the charge like the current one verifies that the right-hand
side of eq.(\ref{anomaly}) coincides with the topological charge density.
In coordinate space this implies $\langle \delta J_x^5 \rangle_A = e Q_x$
which is exactly the anomaly equation on the lattice.

At this point we have shown that a lattice regularization of chiral fermions
is possible. In particular, neutrinos can exist on the lattice. In this respect
the lattice is as good as any other regularization scheme. Although
our calculation was limited to first order perturbation theory, by
construction it can be extended to all orders. In fact, it should be possible
to generalize Reisz' lattice power counting theorems \cite{Rei88} such that
they apply to our fermions. Still, it is not completely clear if our
construction can provide
a nonperturbative definition of chiral gauge theories. As it stands our action
is gauge invariant up to $O(e)$. Before using it in a numerical simulation it
should be made gauge invariant to all orders by hand. This is straightforward.
But when one uses this action in a simulation at large $e$ we cannot exclude
that the chiral properties of the fermions are affected.
Still, when the gauge couplings are weak --- as they are in the Standard model
--- it should be possible to use our action even though other couplings (as the
Yukawa coupling of the top quark) are strong.

Since we started from a perturbatively renormalizable theory in the
continuum we implicitly assumed that all gauge anomalies have been
cancelled. Once we are on the lattice it seems that one could change the
fermion content by hand and make the theory anomalous. However, by doing so
one looses contact with the underlying continuum theory and it is not clear
if the resulting lattice theory makes sense.

We emphasize that our lattice
theory is obtained naturally by integrating out the high momentum modes of the
fields. Hence it contains the same physics as the continuum theory we started
with. Since our action is
nonlocal it remains to be seen if standard numerical techniques can be
applied. We like to point out that the action is perfect in the sense of
Hasenfratz and Niedermayer \cite{Has94}, i.e. there are no cut-off artifacts,
at least to order $e$. Hence it may be sufficient to
work on small lattices. Of course, one could also use the action in vector-like
theories like QCD. However, there are also local perfect actions that are
more practical \cite{Wie93}.
Although they break chiral symmetry the breaking has no effect on the spectrum.
In fact, as a by-product of our investigation we have obtained the perfect
action for a massive fermion coupled to a gauge field to $O(e)$ \cite{Bie95}.

We thank P. Hasenfratz and F. Niedermayer for interesting discussions about
the lattice renormalization group.
We are also grateful to T. Kalkreuter and A. Kronfeld for bringing to our
attention the references \cite{Cro85} and \cite{Wil76}, respectively.

\end{document}